\documentclass[sigconf]{acmart}

\usepackage{enumitem}

\AtBeginDocument{%
  \providecommand\BibTeX{{%
    \normalfont B\kern-0.5em{\scshape i\kern-0.25em b}\kern-0.8em\TeX}}}

\copyrightyear{2022}
\acmYear{2022}
\setcopyright{acmcopyright}\acmConference[CHI '22]{CHI Conference on Human Factors in Computing Systems}{April 29-May 5, 2022}{New Orleans, LA, USA}
\acmBooktitle{CHI Conference on Human Factors in Computing Systems (CHI '22), April 29-May 5, 2022, New Orleans, LA, USA}
\acmPrice{15.00}
\acmDOI{10.1145/3491102.3501977}
\acmISBN{978-1-4503-9157-3/22/04}

\begin{document}

\title{Lattice Menu: A Low-Error Gaze-Based Marking Menu Utilizing Target-Assisted Gaze Gestures on a Lattice of Visual Anchors}

\settopmatter{authorsperrow=4}

\author{Taejun Kim}
\affiliation{%
  \institution{HCI Lab, KAIST}
  \city{Daejeon}
  \country{South Korea}}
\email{taejun.kim@kaist.ac.kr}
\author{Auejin Ham}
\affiliation{%
  \institution{HCI Lab, KAIST}
  \city{Daejeon}
  \country{South Korea}}
\email{hamaj@kaist.ac.kr}
\author{Sunggeun Ahn}
\affiliation{%
  \institution{HCI Lab, KAIST}
  \city{Daejeon}
  \country{South Korea}}
\email{topmaze@kaist.ac.kr}
\author{Geehyuk Lee}
\affiliation{%
  \institution{HCI Lab, KAIST}
  \city{Daejeon}
  \country{South Korea}}
\email{geehyuk@gmail.com}

\renewcommand{\shortauthors}{Kim et al.}

\begin{abstract}

We present Lattice Menu, a gaze-based marking menu utilizing a lattice of visual anchors that helps perform accurate gaze pointing for menu item selection. Users who know the location of the desired item can leverage target-assisted gaze gestures for multilevel item selection by looking at visual anchors over the gaze trajectories. Our evaluation showed that Lattice Menu exhibits a considerably low error rate (\textasciitilde1\%) and a quick menu selection time (1.3-1.6 s) for expert usage across various menu structures (4 $\times$ 4 $\times$ 4 and 6 $\times$ 6 $\times$ 6) and sizes (8, 10 and 12\textdegree{}). In comparison with a traditional gaze-based marking menu that does not utilize visual targets, Lattice Menu showed remarkably (\textasciitilde5 times) fewer menu selection errors for expert usage. In a post-interview, all 12 subjects preferred Lattice Menu, and most subjects (8 out of 12) commented that the provisioning of visual targets facilitated more stable menu selections with reduced eye fatigue.

\end{abstract}

\begin{teaserfigure}
  \centering
  \includegraphics[width=\textwidth]{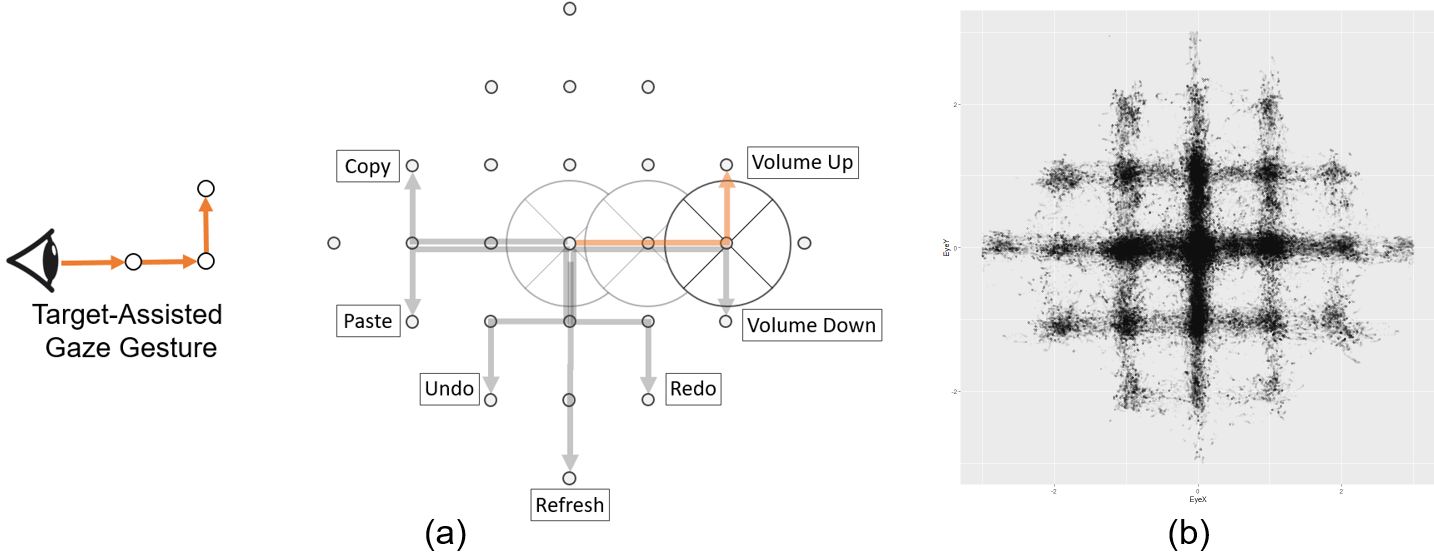}
  \caption{(a) Lattice Menu allows users to perform stable target-assisted gaze gestures by providing a lattice of visual anchors. (b) Distribution of eye saccade landing positions on Lattice Menu collected from User Study 2 (n $\approx$ 300,000).}
  \label{fig:teaser}
  \Description{(a) Lattice Menu allows users to perform stable target-assisted gaze gestures by providing a lattice of visual anchors. (b) Distribution of eye saccade landing positions on Lattice Menu collected from User Study 2 (n $\approx$ 300,000).}
\end{teaserfigure}




\maketitle

\section{Introduction}

Hands-free system control with a gaze-based menu \cite{ahn2021stickypie, urbina2010pies, kammerer2008looking} has become more viable with the advancement of eye tracking technology. In particular, the integration of eye trackers into major augmented reality (AR) and virtual reality (VR) headsets, such as Hololens 2, Vive Pro Eye, and FOVE VR, makes a gaze-based menu more feasible for AR/VR applications.

A marking menu \cite{kurtenbach1991issues, kurtenbach1993design} offers fast selection skills for expert users by drawing a mark in the direction of the desired menu item. Users who already know the location of the desired item in the menu do not have to navigate a hierarchical menu \cite{scarr2011dips, cockburn2014supporting} step by step. The gradual transition from novice to expert is supported as physical actions made by a novice (i.e., navigating the GUI menu for selecting the desired item) become rehearsal of expert gesture skills \cite{kurtenbach1994user}. Transferring these benefits of a marking menu into the gaze input domain, i.e., gaze-based marking menus, is a valuable research goal. 

Border-crossing selection \cite{urbina2010pies} (Figure \ref{fig:userstudy2explain}a) is a first step for realizing gaze-based marking menus. It allows users to select a menu item by crossing the border of a pie slice with their eye movements. Expert users can perform a fast selection skill with successive border-crossing eye movements (i.e., zig-zag gaze gestures) that mimic the hand-drawn mark of the original marking menu. A gradual transition from novice to expert is also supported because users can memorize the required gaze gesture path up to the target command over time. Urbina et al. \cite{urbina2010pies} reported that the border-crossing selection method produced a better performance than a dwell-based method.

However, using gaze gestures ``in the air'' (i.e., border-crossing \cite{urbina2010pies}) for a marking menu can be error-prone because there are several unwanted options located adjacent to a single wanted item in a menu structure. In previous studies on eye gaze gestures \cite{majaranta2019inducing, hornof2003eyedraw, wobbrock2008longitudinal, drewes2007interacting, hyrskykari2012gaze, porta2008eye}, researchers have proposed the use of visual targets to anchor the gaze for accurate eye movements. EyeDraw \cite{hornof2003eyedraw} uses a grid of dots and was reported to be helpful for eye-drawing tasks. In the EyeWrite interface \cite{wobbrock2008longitudinal}, small visual anchors are placed at the four corners of the input box to support precise eye-typing movements. Inspired by previous studies \cite{majaranta2019inducing, hornof2003eyedraw, wobbrock2008longitudinal, hyrskykari2012gaze}, we attempted to integrate visual anchors into the gaze-based marking menu technique to facilitate stable eye control. 

Herein, we present Lattice Menu, a gaze-based marking menu with a lattice of visual anchors for supporting accurate gaze pointing. As depicted in Figure \ref{fig:latticeMenuUsage}, users can select a menu item by looking at the corresponding visual anchor outside the border of the pie slice. Experienced users, who already know the location of the desired item in the menu, can perform target-assisted gaze gestures over the lattice of visual anchors. In this study, we implemented and evaluated Lattice Menu technique in a VR environment.

In the remainder of this paper, we review related works, describe the design of Lattice Menu, and explain the overall structure of the user study. Preliminary studies are first described to demonstrate the process of optimizing the Lattice Menu technique. Subsequently, we evaluate the performance of the designed Lattice Menu on various menu layouts (tested structures, 4 $\times$ 4 $\times$ 4 and 6 $\times$ 6 $\times$ 6; menu radius, 8, 10, and 12\textdegree{}) in User Study 1. Finally, we compare the performance of Lattice Menu with the traditional border-crossing marking menu \cite{urbina2010pies} that does not utilize visual targets in User Study 2.

The main contributions of this study are as follows:
\begin{itemize}[topsep=1pt, partopsep=1pt]
  \item We introduce Lattice Menu, a gaze-based marking menu technique that supports target-assisted gaze gestures on a lattice of visual anchors.
  \item Through empirical evaluation, we show that Lattice Menu provides a stable user control with a remarkably low error rate (\textasciitilde1\%).
\end{itemize}

\section{Related Work}

\subsection{Marking Menu Technique}

A marking menu \cite{kurtenbach1993design}, initially proposed in the hand-based input domain (e.g., stylus and mouse), is a radial menu that allows users to make a selection by drawing a mark to the desired item. The original marking menu offered two different modes for selection; novice users wait for a certain amount of time after pressing the input device to trigger the display of the GUI menu (\textit{menu mode}), whereas expert users simply draw a mark to the desired item without waiting (\textit{mark mode}). The marking input is scale-invariant through the post-processing of gesture decoding. 

Meanwhile, Henderson et al. \cite{henderson2020investigating} recently questioned the necessity of the delayed display and mode separation. They failed to reveal the obvious performance advantage compared to the immediately displayed design without a mode separation. We are now curious about the essential aspects of a marking menu.
%Now, one can say, what are the essential requirements to be ``defined'' as a marking menu? 

We believe that there are two fundamental benefits in a marking menu; (1) support for faster menu selection through a gestural input, i.e., a mark, and (2) the principle of rehearsal \cite{kurtenbach1993design}, i.e., a gradual transition from a novice to an expert. We believe that the separate modes or the scale-invariance may be just a means to achieve these goals.

Lattice Menu does not separate modes for selection, nor does it have scale-invariant features. However, we call Lattice Menu a marking menu because it supports (1) expert skills through a gestural input (i.e., target-assisted gaze gesture) and (2) the principle of rehearsal; eye movements for novice users' selection become rehearsals for expert skill. Lattice Menu is a new marking menu designed for the eye gaze input domain.

\subsection{Gaze-Based Marking Menu Techniques}

\begin{figure}[t]
  \centering
  \includegraphics[width=8cm]{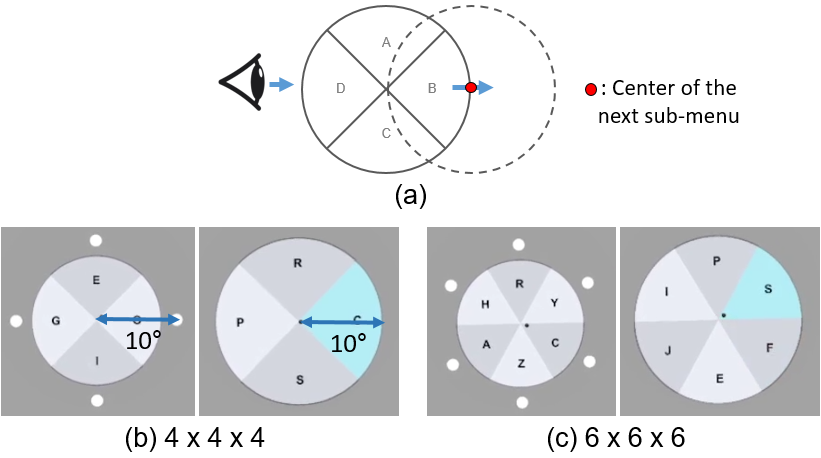}
  \caption{(a) Border-crossing marking menu (\textit{BorderPie}, re-illustrated from \cite{urbina2010pies}). Example layouts of \textit{Lattice Menu} and \textit{BorderPie} of 10\textdegree{} \textit{Size} with (b) 4 $\times$ 4 $\times$ 4 and (c) 6 $\times$ 6 $\times$ 6 \textit{Structure}.}
  \label{fig:userstudy2explain}
  \Description{a) Border-crossing marking menu illustration. Example layouts of Lattice Menu and BorderPie with s10 degree radius}
\end{figure}

Despite their potential benefits, only a few studies \cite{urbina2010pies, huckauf2008gazing, ahn2021stickypie} have explored gaze-based marking menu designs. Border-crossing selection \cite{urbina2010pies} was the first step for a gaze-based marking menu technique. The border-crossing marking menu allows users to select a menu item by crossing the border of the pie slice with eye movements (Figure \ref{fig:userstudy2explain}a). Expert skill with successive border-crossing eye movements, i.e., gaze gestures, can be supported. This approach was also reported to be effective for preventing an unintentional selection compared to dwell-based menu selection \cite{majaranta2009fast, tien2008improving}. 

Huckauf et al. \cite{huckauf2008gazing} introduced the \textit{pEYE} depicted in Figure \ref{fig:comparisonWithpEYE}b, which employs a pizza-crust-like area for gaze pointing guidance. Their study on \textit{pEYE} did not focus on an investigation of the marking menu technique (e.g., expert-level hierarchy); however, the idea of visual guidance for stable eye control is valuable. We compare the performance of the \textit{pEYE} design with our approach in Section 3.4.

StickyPie \cite{ahn2021stickypie} was recently proposed to overcome the overshooting problem (i.e., unwanted menu selection owing to the overshot border-crossing eye movement) of the traditional technique by locating the sub-menu on the estimated saccade landing position. An evaluation of StickyPie showed a faster menu selection time and a lower error rate than the traditional technique \cite{urbina2010pies}. We considered performance comparison of ours with StickyPie, but we had a practical difficulty. Careful setting of parameters is crucial for the optimal performance of StickyPie, but the parameters used in the published paper could not be adopted because the interaction context of the StickyPie differs from that of ours (e.g., eyes-only input vs. head-combined gaze input). As our focus in the current study is to validate the effectiveness of the provisioning of visual targets, we decided to compare ours with the traditional border-crossing marking menu \cite{urbina2010pies} in this study. Instead, we compared the reported performances of StickyPie with ours in Section 8.4.

\subsection{Target-Assisted Gaze Gesture}

\begin{figure}[t]
  \centering
  \includegraphics[width=8cm]{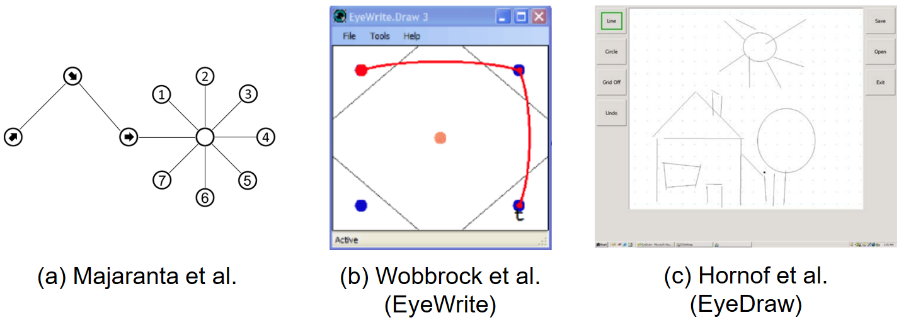}
  \caption{Previous studies that adopted visual targets for stable gaze gestures. (a)  \cite{majaranta2019inducing}, (b) \cite{wobbrock2008longitudinal}, and (c) \cite{hornof2003eyedraw}}
  \label{fig:visualtargets}
  \Description{Previous studies that adopted visual targets for stable gaze gestures. (a)  Majaranta et al.'s study, (b) wobbrock et al.'s study, and (c) Hornof et al.'s study}
\end{figure}

Majaranta et al. \cite{majaranta2019inducing} noted that \textit{"Having something to look at not only guides the gaze but also avoids the need to perform unnatural, potentially uncomfortable gaze gestures"}. Previous studies \cite{hornof2003eyedraw, wobbrock2008longitudinal, drewes2007interacting, hyrskykari2012gaze, porta2008eye} have employed visual targets for achieving stable gaze gestures, i.e., target-assisted gaze gestures.  Drewes et al. \cite{drewes2007interacting} compared three gaze gestures with and without visual targets and reported that many subjects failed to perform complex gaze gestures without the help of visual targets. As shown in Figure \ref{fig:visualtargets}b, EyeWrite \cite{wobbrock2008longitudinal} utilized small point targets in the four corners to support stable gaze gestures for text entry. EyeDraw \cite{hornof2003eyedraw} shown in Figure \ref{fig:visualtargets}c utilized a grid of dots for eye-drawing tasks and reported that it was helpful to EyeDraw users. 

Inspired by previous studies, Lattice Menu was designed to allow users to perform target-assisted gaze gestures over the lattice of visual anchors. Compared to gaze gestures ``in the air'' without any target to look at \cite{isomoto2020gaze}, users can achieve more stable menu selections. In the preliminary studies of this paper, we further explore several design dimensions of visual anchors to better support the target-assisted gaze gestures.

\section{Lattice Menu}

\begin{figure*}[t]
  \centering
  \includegraphics[width=\textwidth]{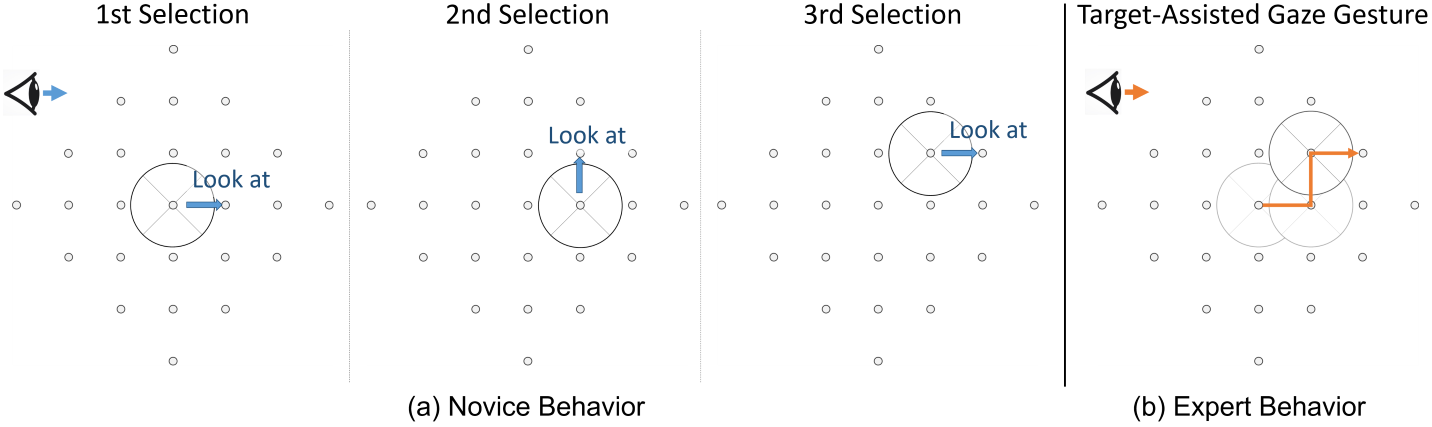}
  \caption{Illustration of (a) novice user behavior and (b) experienced user behavior of Lattice Menu. }
  \label{fig:latticeMenuUsage}
  \Description{Illustration of (a) novice user behavior and (b) experienced user behavior of Lattice Menu. Novice user navigates the menu to find and select the desired item. Experienced user performs target-assisted gaze gestures over the lattice of visual anchors}
\end{figure*}

Lattice Menu is a gaze-based marking menu with a lattice of visual anchors that helps accurate gaze pointing. Users can select a menu item by looking at the corresponding visual target (i.e., visual anchor) outside the border of the pie slice. When the eye gaze enters the \textit{Item Selection Zone} (Figure \ref{fig:latticeMenuParameters}a) around the visual anchor, a corresponding menu item is immediately selected. Experienced users, who already know the required trajectory to the desired item, can perform target-assisted gaze gestures over the lattice of visual anchors as depicted in Figure \ref{fig:latticeMenuUsage}b. In this study, we implemented and evaluated the Lattice Menu technique in a VR environment, under a three-level hierarchical menu structure, covering 64 commands for a 4  $\times$ 4  $\times$ 4 menu structure, and 216 commands for a 6  $\times$ 6  $\times$ 6 menu structure. 

\subsection{Head-Fixed Menu vs. World-Fixed Menu in AR/VR}

There can be two types of menus in the AR/VR environment: (1) a menu that is fixed at the headset display, \textit{Head-Fixed Menu} (i.e., a menu that follows users' head movement) or (2) a menu located in a fixed coordinate of the AR/VR world, \textit{World-Fixed Menu}. An eyes-only input is used for the \textit{Head-Fixed Menu}, and a head-combined gaze input is used for the \textit{World-Fixed Menu}. 

Validation in the \textit{World-Fixed Menu} setup has the advantage of being compatible with an external, stationary eye tracker (e.g., for smart TV) environment, because a head-combined gaze input is utilized in both cases. By contrast, validation in the eyes-only input (\textit{Head-Fixed Menu}) setup also has its advantage, being the only possible option when 3D sensing for the surrounding space is incapable of (e.g., smart glasses without 3D sensing capability). In this study, we decided to validate our Lattice Menu technique in the \textit{World-Fixed Menu} setup.

\subsection{Design Rationale}

\begin{figure}[t]
  \centering
  \includegraphics[width=8.5cm]{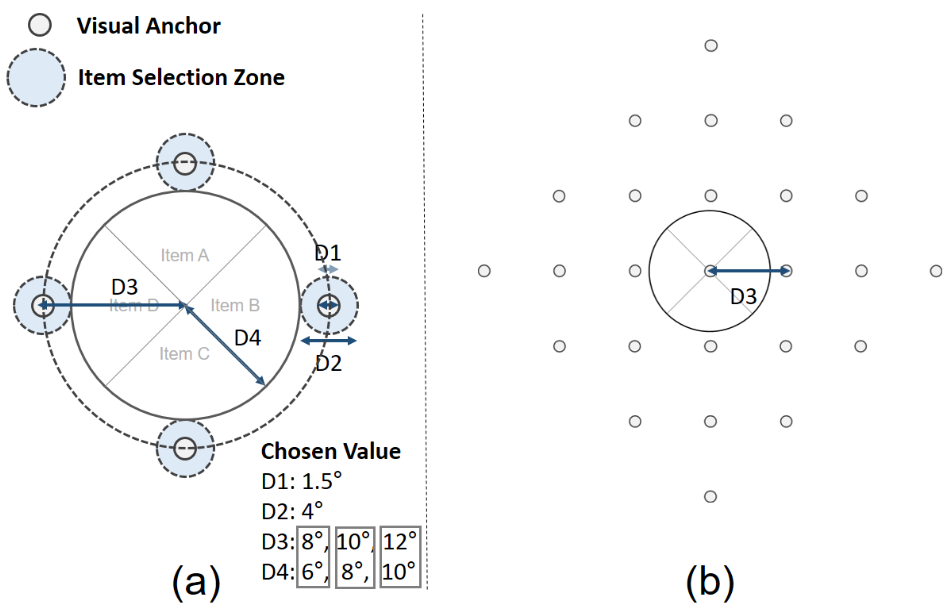}
  \caption{(a) Controllable parameters of Lattice Menu. (b) 4 $\times$ 4 $\times$ 4 Lattice Menu. D1, width of visual anchor; D2, width of \textit{Item Selection Zone}; D3, effective radius of Lattice Menu, and D4, radius of visual pie.}
  \label{fig:latticeMenuParameters}
  \Description{(a) Controllable parameters of Lattice Menu. (b) 4 $\times$ 4 $\times$ 4 Lattice Menu. D1, width of visual anchor; D2, width of Item Selection Zone; D3, effective radius of Lattice Menu, and D4, radius of visual pie.}
\end{figure}

The controllable parameters of Lattice Menu and our chosen values are shown in Figure \ref{fig:latticeMenuParameters}a. In this section, we describe the rationale of our design. First, we decided to use a circular-type visual anchor, which is commonly applied in gaze pointing tasks \cite{hornof2003eyedraw, wobbrock2008longitudinal, thaler2013best, majaranta2019inducing, hyrskykari2012gaze}. We decided to set \textit{D1} (the width of the visual anchor) to 1.5\textdegree{} referring to the fixation target size adopted in previous vision studies \cite{thaler2013best}. In addition, \textit{D2} (the width of the \textit{Item Selection Zone}) was determined to be 4\textdegree{} considering the advertised tracking accuracies of major VR headsets \cite{FOVE, rajanna2018gaze, ViveProEye}, which are better than 2\textdegree{}; the advertised tracking accuracy of the FOVE \cite{FOVE} is 1.15\textdegree{}, and that of HTC Vive Pro Eye \cite{ViveProEye} is 0.5-1.1\textdegree{}. 

For the menu size (\textit{D3}: the effective radius of Lattice Menu), we used 10\textdegree{} as the base size, allowing a single-level menu selection to be performed with a comfortable eye-only movement (i.e., within a range of $\pm$ 15\textdegree{} \cite{menozzi1994direction}), and multi-level target-assisted gaze gestures to be performed with a comfortable head-combined eye movement \cite{sidenmark2019eye}. Preliminary studies were conducted with the base size (10\textdegree{}). In User Studies 1 and 2, three menu sizes, i.e., 8\textdegree{} (small), 10\textdegree{} (base), and 12\textdegree{} (large), were tested for a broader evaluation.

Finally, for \textit{D4} (the radius of a visual pie), the need for this parameter has to be explained first. The reason we do not simply set \textit{D4} same with D3 (i.e., the reason we make visual pie smaller than the effective radius of Lattice Menu) is to separate the area of the visual pie from the \textit{Item Selection Zone} as shown in Figure \ref{fig:latticeMenuParameters}a. In an informal pilot study, unintentional selection errors frequently occurred while reading item labels when D4 was the same as D3. However, after we separated the visual pie from the \textit{Item Selection Area}, the errors were effectively reduced. This observation implies that the visual pie can ``capture'' eye gaze of the user inside the boundary, which is consistent with the previous finding that the ``eyes prefer to stay within the same object'' \cite{theeuwes2010object}.

\subsection{Distance from Content to Item Selection Zone}

\begin{figure}[b]
  \centering
  \includegraphics[width=8cm]{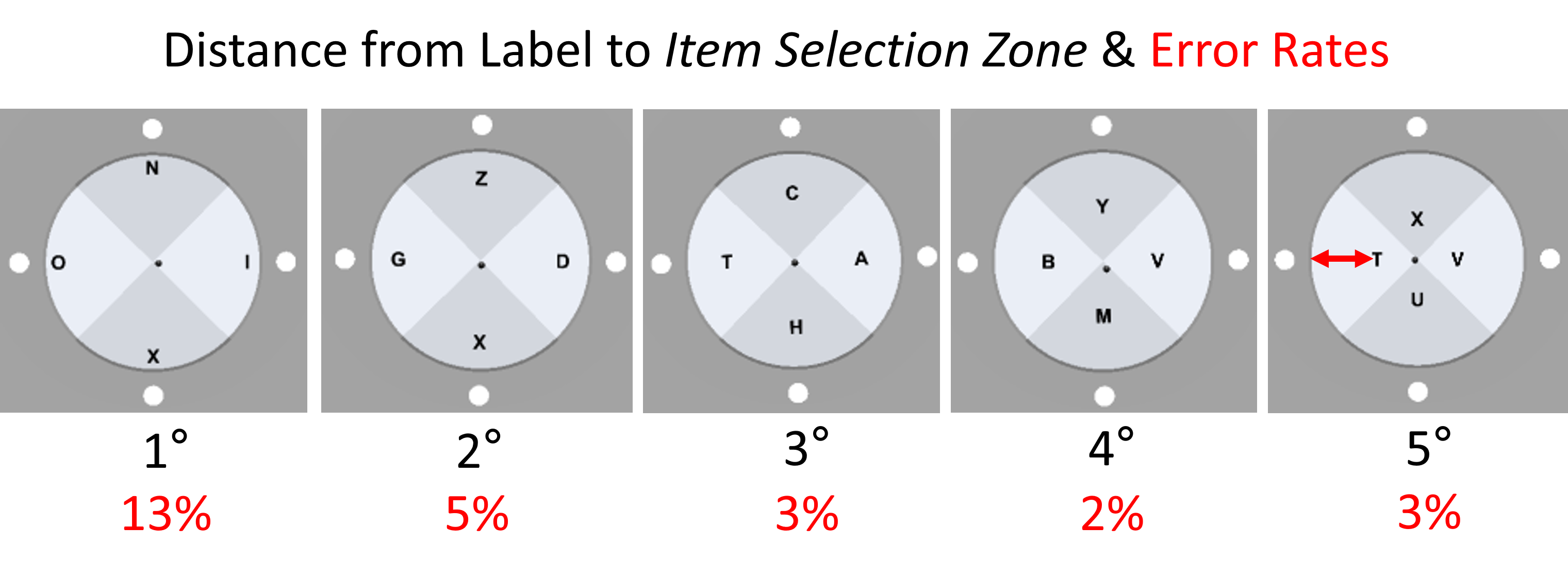}
  \caption{Distance from item label to \textit{Item Selection Zone} and the observed error rates for each condition.}
  \label{fig:distanceToSelectionZone}
  \Description{Distance from item label to Item Selection Zone and the observed error rates for each condition. Error rates were 13, 5, 3, 2, 3 percentages for each distance condition 1, 2, 3, 4, 5 degree}
\end{figure}

If the menu item contents (e.g., label texts) are located too close to the \textit{Item Selection Zone}, unintentional selection errors can easily occur during a visual search of the items. For the contents placement guidelines, we conducted a pilot study (n = 10) to examine the effect of the distance from the menu item content to the \textit{Item Selection Zone} on the error rates. We collected 540 \textit{Novice Trials} (described in Section 4.3) for each 1\textdegree{} to 5\textdegree{} \textit{Distance} conditions under 4 $\times$ 4 $\times$ 4 menu structure with the base menu size (10\textdegree{}). The order of the conditions across subjects was counterbalanced using a Balanced Latin Square. The results are shown in Figure \ref{fig:distanceToSelectionZone} and a Friedman test revealed that the effect of the \textit{Distance} on the error rates was significant ($\chi^2$(4) = 15.653, \textit{p} < .005). The results imply that a \textit{Distance} of 3\textdegree{} or more is required to prevent unintentional errors during the visual search of the menu contents, which is consistent with the previous gaze-based menu guidelines \cite{ahn2021stickypie}. From this observation, we fixed the \textit{Distance} to 3\textdegree{} in the remaining studies.

\subsection{Comparison with pEYE}

\begin{figure}[b]
  \centering
  \includegraphics[width=6cm]{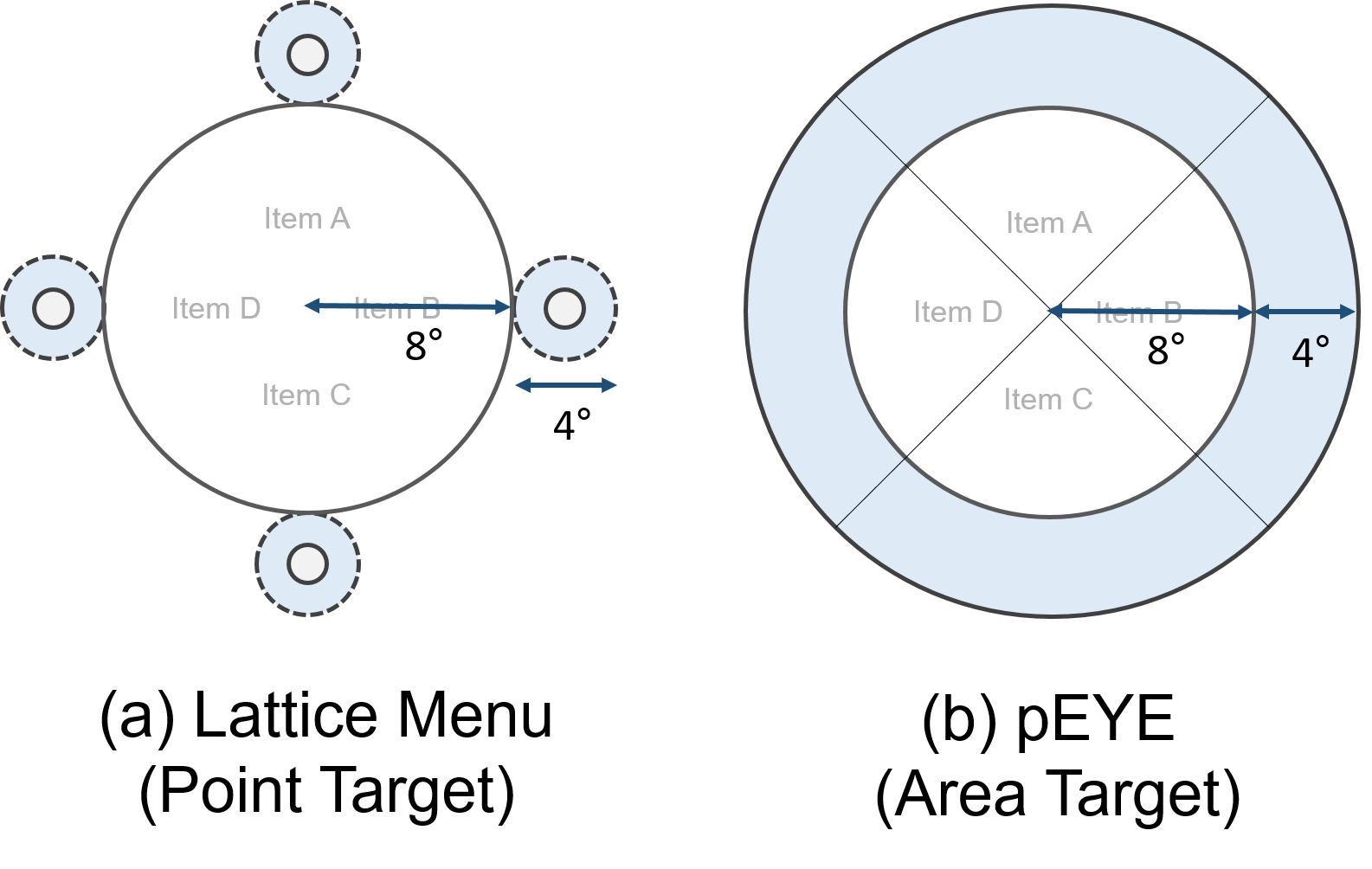}
  \caption{Areas for item selection in (a) Lattice Menu and (b) \textit{pEYE} \cite{huckauf2008gazing} }
  \label{fig:comparisonWithpEYE}
  \Description{Areas for item selection in (a) Lattice Menu and (b) pEYE}
\end{figure}

Our design was inspired by \textit{pEYE} \cite{huckauf2008gazing} shown in Figure \ref{fig:comparisonWithpEYE}b. The \textit{pEYE} employed a pizza-crust-like area outside the border for gaze pointing guidance. We decided to compare the performance of the two designs through a pilot study. We expected that a more localized point target would better support a stable gaze control than the pizza-crust-like target of \textit{pEYE}.

To validate our expectations, we conducted a 2-day in-lab pilot study (n = 6). We collected 120 \textit{Novice Trials} and 240 \textit{Experienced Trials}  (described in Section 4.3) of the menu selection task for each technique under the menu layout depicted in Figure \ref{fig:comparisonWithpEYE}. The order of the conditions was counterbalanced across the subjects. As a result, we observed generally higher error rates in \textit{pEYE}. Particularly for \textit{Experienced Trials}, \textit{pEYE} showed error rates six-times higher than Lattice Menu (1.1\ vs. 6.1\%), which validates our initial expectation and the superiority of the Lattice Menu design.

\begin{figure*}[t]
  \centering
  \includegraphics[width=0.85\textwidth]{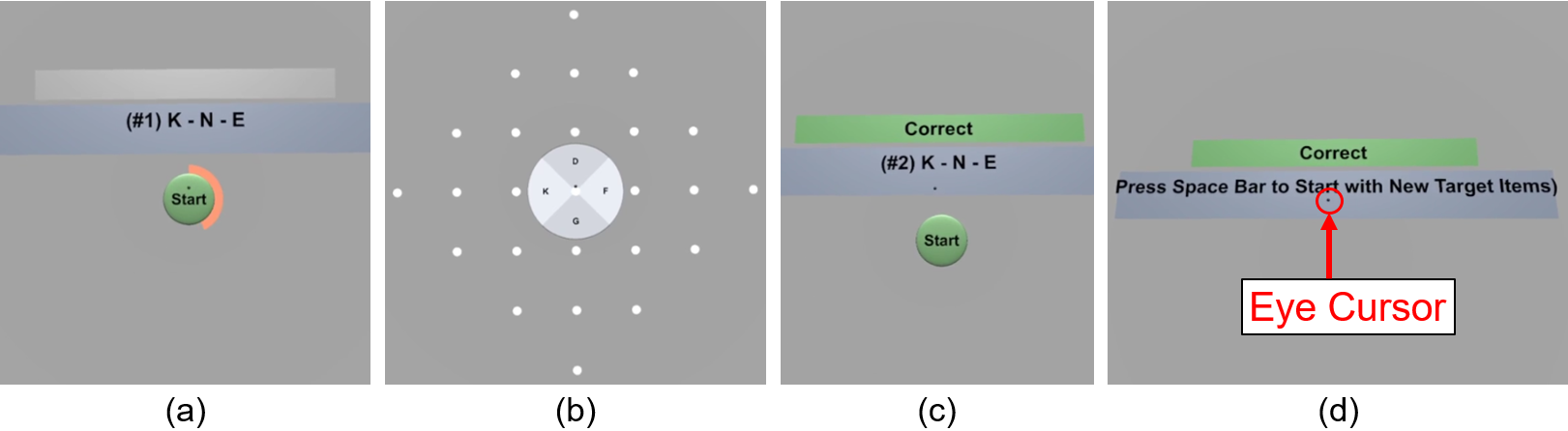}
  \caption{Sequence of user tasks in VR testing program. (a) After memorizing given item sequence (i.e., K-N-E), subjects dwell over start button, and (b) select items using Lattice Menu. (c) Results are shown after each trial (i.e., correct). (d) After repeating each sequence four times, subjects pressed space bar to obtain new target items.}
  \label{fig:VRenvironment}
  \Description{Sequence of user tasks in VR testing program. (a) After memorizing given item sequence (i.e., K-N-E), subjects dwell over start button, and (b) select items using Lattice Menu. (c) Results are shown after each trial (i.e., correct). (d) After repeating each sequence four times, subjects pressed space bar to obtain new target items.}
\end{figure*}

\section{User Study Overview}

In preliminary studies, we explored several design dimensions of visual anchors to optimize and finalize our Lattice Menu design. In User Study 1, we evaluated the designed Lattice Menu across various menu layouts (tested structures of 4 $\times$ 4 $\times$ 4 and 6 $\times$ 6 $\times$ 6, and tested radii of 8\textdegree{}, 10\textdegree{}, and 12\textdegree{}). In User Study 2, we compare the performance of Lattice Menu with the traditional border-crossing marking menu \cite{urbina2010pies} that does not utilize visual targets.

\subsection{Participants and Participant Screening Process}

A group of 12 subjects (4 females, mean age of 24.0 and SD of 5.3) participated in preliminary study 1 (days 1 and 2), preliminary study 2 (day 3), and User Study 1 (day 4) on consecutive days. Afterward, for a fair comparative study, a new group of 12 subjects (three females, mean age of 22.0 and SD of 2.0) with no prior experience of Lattice Menu was recruited for User Study 2. Each subject was paid 15,000 KRW ($\approx$ 13 USD) per hour as a reward.

The width of the \textit{Item Selection Zone}, as shown in Figure \ref{fig:latticeMenuParameters}, was chosen at 4\textdegree{}. This decision was based on the advertised tracking accuracies of major VR headsets, which are better than 2\textdegree{}; the advertised tracking accuracy of the FOVE \cite{FOVE} is 1.15\textdegree{}, and that of HTC Vive Pro Eye \cite{ViveProEye} is 0.5-1.1\textdegree{}. However, FOVE that we used in this study exhibited tracking errors larger than 2\textdegree{} for some users. Therefore, we had to screen two and four applicants from the first and second groups, respectively, who repeatedly showed tracking errors larger than 2\textdegree{} in a 9-point accuracy test we developed.

\subsection{Common Apparatus}

Lattice Menu was implemented in a VR environment using Unity. FOVE \cite{FOVE}, a VR headset with an integrated eye tracker was used. FOVE has a display resolution of 2560 $\times$ 1440 pixels and a field of view of 100\textdegree{} with a frame rate of 70fps. The built-in binocular eye tracking system of FOVE has a sampling rate of up to 120 Hz.

The ray casting of the eye gaze was computed using the gaze origin and gaze vector collected from each eye. The 2D gaze positions hit on the rendered menu plane were computed using the gaze rays from both eyes. A combined gaze position was calculated by averaging left/right 2D gaze positions. The eye cursor indicating combined gaze position was visualized with a dark-gray circle (diameter of 0.5\textdegree{}) as shown in Figure \ref{fig:VRenvironment}d.

\subsection{Common Task and VR environment}

\begin{figure}[b]
  \centering
  \includegraphics[width=8cm]{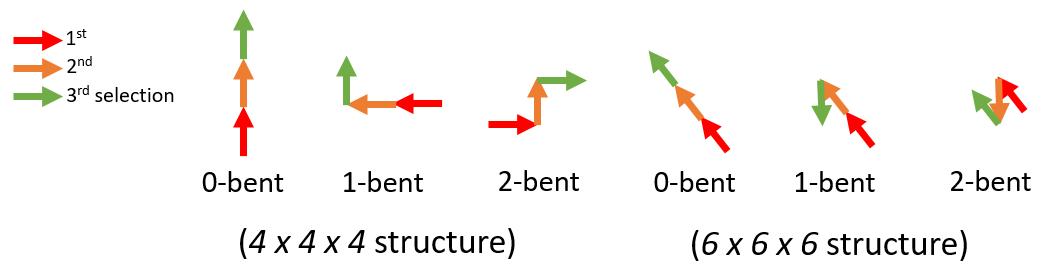}
  \caption{Different types of target item paths used in our studies. }
  \label{fig:taskPaths}
  \Description{Different types of target item paths used in our studies.}
\end{figure}

Figure \ref{fig:VRenvironment} shows how subjects performed the task using the VR testing program. First, three menu items (e.g., K-N-E) to be selected in each menu level are shown. After memorizing the target items, subjects can stare at the start button for 1 s to start a task trial. After the menu selection is made, the result of the trial is provided, as shown in Figure \ref{fig:VRenvironment}c (i.e., correct or incorrect)

The same target items (e.g., K-N-E) were repeated four times to simulate an expert user performance. In the first trial of repetition, subjects should navigate the GUI menu to find the target items. In the latter trials of repetition, subjects were able to follow the known trajectory up to the target item they found before. They were informed about the strategy of the target-assisted gaze gesture, consecutively looking at visual anchors along the way. After four repetitions with the same target items, subjects could obtain a new target item by pressing the space bar (Figure \ref{fig:VRenvironment}d). When the space bar was pressed, new randomized target items were provided to subjects. The item (i.e., letters) arrangement of the menu was also randomized. 

In this study, we denote the first (\#1) trials of repetition as \textit{Novice Trials}, and the third and fourth (\#3 and \#4) trials as \textit{Experienced Trials}. To test a range of menu positions, we randomly chose an appropriate number of target item paths from \textit{0-bent}, \textit{1-bent}, and \textit{2-bent}, depicted in Figure \ref{fig:taskPaths}, for testing.

\subsection{Common Procedures}

Subjects sat on a chair and were instructed on how to conduct the task through a demonstration video before starting the experiment. After putting the headset, the eye tracker was calibrated using the default calibration method of FOVE, and subjects could request a recalibration when they felt a mismatch with the position of the eye cursor. They were instructed to perform the task as accurately and quickly as possible and were informed that they could use eye and head movements together. Subjects took a short break between each block or experimental condition to relieve eye fatigue.

\section{Preliminary Study}

Preliminary user studies were conducted to finalize our Lattice Menu design. Here, we explore several design possibilities that can significantly affect the usability of Lattice Menu. Experiments were conducted under the baseline menu layout (menu structure of 4 $\times$ 4 $\times$ 4, effective radius (\textit{D3}) of 10\textdegree{}, and visual pie radius (\textit{D4}) of 8\textdegree{}).

\subsection{Preliminary Study 1: Progressive Unfolding Effect}

\begin{figure}[t]
  \centering
  \includegraphics[width=8cm]{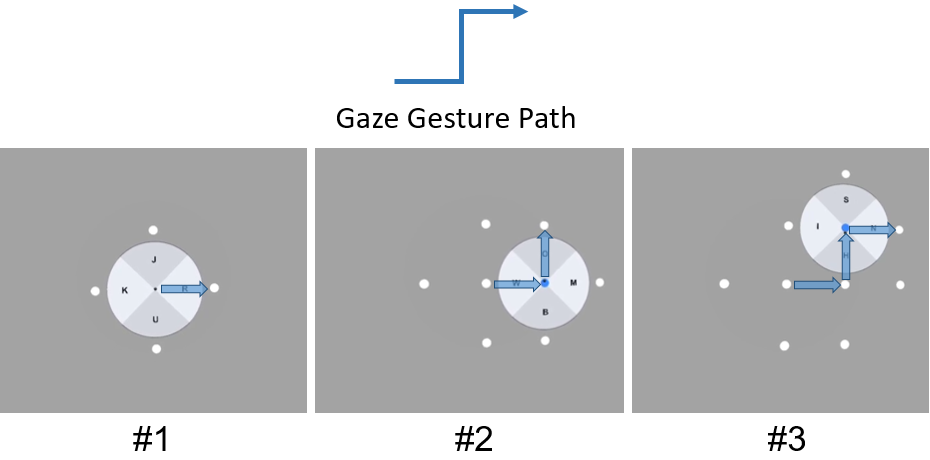}
  \caption{Lattice Menu with \textit{Progressive Unfolding} effect; \textit{Progressive Lattice Menu} }
  \label{fig:progressiveUnfoldingEffect}
  \Description{Lattice Menu with Progressive Unfolding effect}
\end{figure}

Visual distraction owing to the lattice of visual anchors on the screen is a critical usability problem of Lattice Menu. As a solution, we devised the \textit{Progressive Unfolding} effect. When the effect is activated, Lattice Menu does not show all lattices but progressively unfolds the visual anchors used for the item selection of the current menu level. For instance, as shown in Figure \ref{fig:progressiveUnfoldingEffect}, four visual anchors at menu level 1 are visible from the beginning, and, when a user looks at one visual anchor to make a selection, another four visual anchors immediately appear.

In preliminary study 1, we examined the effect of the \textit{Progressive Unfolding Effect} on the user performance. We compare the performance of Lattice Menu with and without the \textit{Progressive Unfolding Effect} (\textit{Progressive Lattice Menu} vs. \textit{Full Lattice Menu}). The experiment was a one-way within-subject design with the following independent variable and levels. The order of the conditions was counterbalanced.

\begin{itemize}
  \item MenuType: \textit{Progressive Lattice Menu}, \textit{Full Lattice Menu}
\end{itemize}

\subsubsection{Procedure}

For each \textit{MenuType}, the same six target paths (two each from \textit{0-bent}, \textit{1-bent}, and \textit{2-bent}, depicted in Figure \ref{fig:taskPaths} chosen at random) were tested. One block consisted of 48 trials (2 \textit{MenuType} $\times$ 6 paths $\times$ 4 repetitions) of the menu selection task and a subject applied three blocks on the first day for the interface training, and five blocks on the second day for testing. It took less than 1 h to complete the experiment each day.

\subsubsection{Analysis}

We collected 720 \textit{Novice Trials} (2 \textit{MenuType} $\times$ 6 paths $\times$ 5 blocks $\times$ 12 subjects) and 1440 \textit{Experienced Trials}, and calculated the task completion time (CT) and error rate (ER). CT was measured from the moment the menu was opened, by staring at the start button for 1 s, until the leaf menu item was selected. A trial was counted as correct only when all three target items were correctly selected. Otherwise, it was counted as an error. For the analysis, we performed a one-way ANOVA on CT and a Friedman test for ER. Because the CT in \textit{Experienced Trials} violated the normality assumption, we performed a Friedman test in this case. 

\subsubsection{Result}

\begin{figure}[t]
  \centering
  \includegraphics[width=8cm]{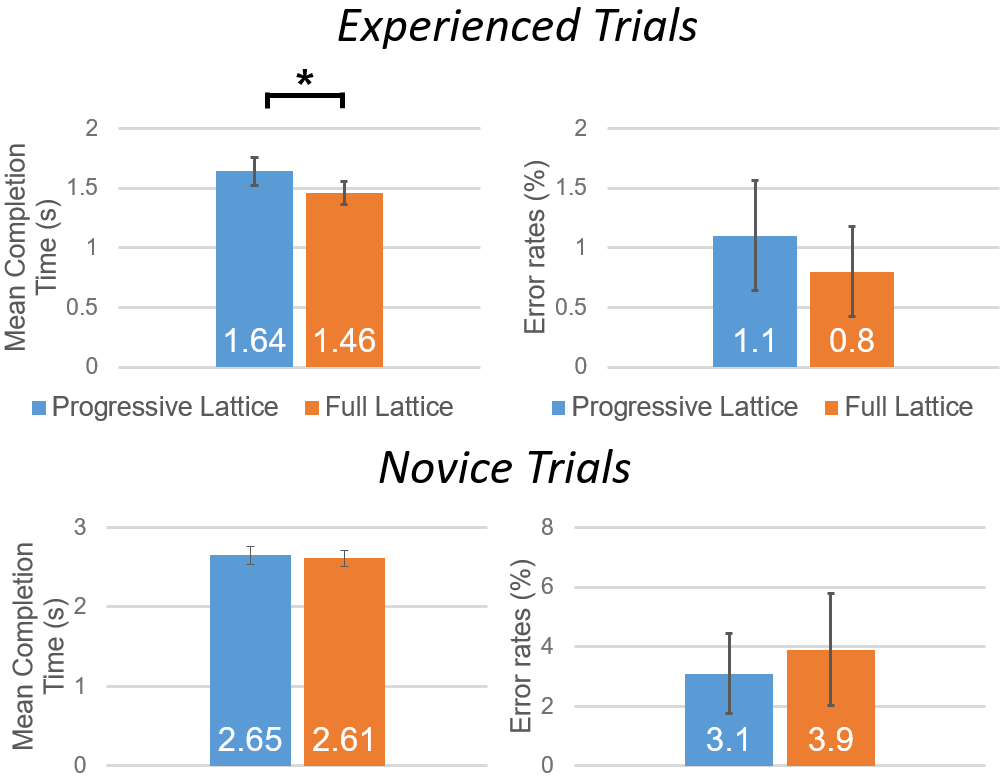}
  \caption{Mean of CT (s) and ER (\%) from preliminary study 1. Error bars show standard errors, and asterisks indicate significant differences (\textit{p} < .005).}
  \label{fig:preliminary1result}
  \Description{Mean of CT and ER from preliminary study 1. Error bars show standard errors, and asterisks indicate significant differences.}
\end{figure}

For the \textit{Experienced Trials}, the effect of \textit{MenuType} was significant on CT ($\chi^2$(1) = 8.333, \textit{p} < .005) but not on ER. The \textit{Full Lattice Menu} showed a significantly faster CT than \textit{Progressive Lattice Menu}. For the \textit{Novice Trials}, the effects of \textit{MenuType} on both CT and ER were not significant.

\subsubsection{Post-interview on user experience}

Despite the significantly longer CT observed in \textit{Progressive Lattice Menu} for \textit{Experienced Trials}, more subjects (four) preferred \textit{Progressive Lattice Menu} than \textit{Full Lattice Menu} (two). Subjects who preferred \textit{Progressive Lattice Menu} mainly commented on the discomfort from the visual distraction experienced in \textit{Full Lattice Menu}. Subject p2 said \textit{"Showing only the necessary anchors makes it easier to maintain my concentration"}. However, subjects who preferred \textit{Full Lattice Menu} mainly commented on the ease of eye movement planning with visual anchors fully shown for up to the last level. However, importantly, most subjects (six) had no particular preference. They commented that they did not feel much difference in terms of the selection performance once they became familiar with both \textit{MenuTypes}.

We had a long discussion on the tradeoff between the advantage of reducing visual distraction and the disadvantage of increasing the menu selection time (0.18s in average) for expert usage that \textit{Progressive Lattice Menu} brings. Finally, we chose to use \textit{Progressive Lattice Menu} for our final design as we judged that the visual distraction of \textit{Full Lattice Menu} can have a more crucial impact on usability, particularly in a realistic environment with different visual contents in the background.

\begin{figure}[t]
  \centering
  \includegraphics[width=5cm]{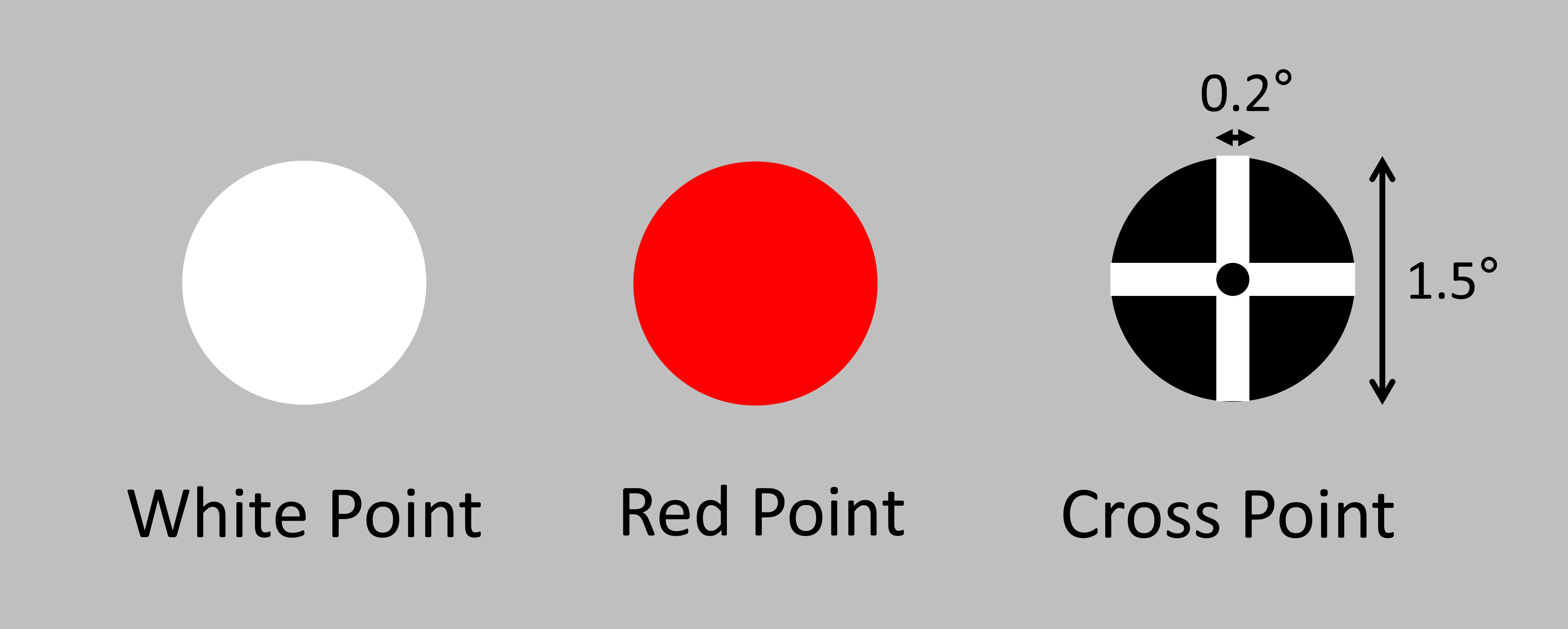}
  \caption{Three types of visual anchors used in preliminary study 2. }
  \label{fig:salientVisualAnchors}
  \Description{Three types of visual anchors used in preliminary study 2. White, Red, and Cross Point targets.}
\end{figure}

\subsection{Preliminary Study 2: Visual Anchor Design - Visual Saliency}

A salient visual target encourages an effortless shift of attention \cite{le2015saccadic, theeuwes1998our, yantis1994stimulus}, and can help with a rapid and accurate eye gaze pointing \cite{deuble1984evaluation, kalesnykas1994retinal, bell2006stimulus}. The visual saliency of the anchors can affect the user experience and performance of Lattice Menu. Herein, we compare three different types of visual anchors: \textit{White Point} (baseline), \textit{Red Point}, and \textit{Cross Point} depicted in Figure \ref{fig:salientVisualAnchors}. \textit{Red Point} is a target with a higher visual saliency than \textit{White Point} in terms of the \textit{Color Opponency} \cite{itti2001computational}. \textit{Cross Point} is another salient target with a high spatial frequency, and has often been used in previous eye-fixation studies \cite{thaler2013best, schuetz2019explanation, bhattarai2019fixation}. 

In preliminary study 2, we investigated the effect of \textit{VisualAnchorType} on user experience and performance of Lattice Menu. The experiment was a one-way within-subject design with the following independent variable and levels. The order of the conditions was counterbalanced.

\begin{itemize}
  \item VisualAnchorType: \textit{White Point}, \textit{Red Point}, \textit{Cross Point}
\end{itemize}

\subsubsection{Procedure}

For each \textit{VisualAnchorType}, the same six target paths (two each from \textit{0-bent}, \textit{1-bent}, and \textit{2-bent} chosen at random) were tested. One block consisted of 72 trials (3 \textit{VisualAnchorType} $\times$ 6 paths $\times$ 4 repetitions) of the menu selection task, and the subject performed five blocks without training because they were already trained after 2 days of experience. It took approximately 1 h to complete the experiment.

\subsubsection{Analysis}

We collected 1080 \textit{Novice Trials} (3 \textit{VisualAnchorType} $\times$ 6 paths $\times$ 5 blocks $\times$ 12 subjects) and 2160 \textit{Experienced Trials}, and calculated the CT and ER. For the analysis, because CT in both \textit{Experienced Trials} and \textit{Novice Trials} violated the normality assumption, we performed Friedman tests on both CT and ER.

\subsubsection{Result}

\begin{figure}[t]
  \centering
  \includegraphics[width=8cm]{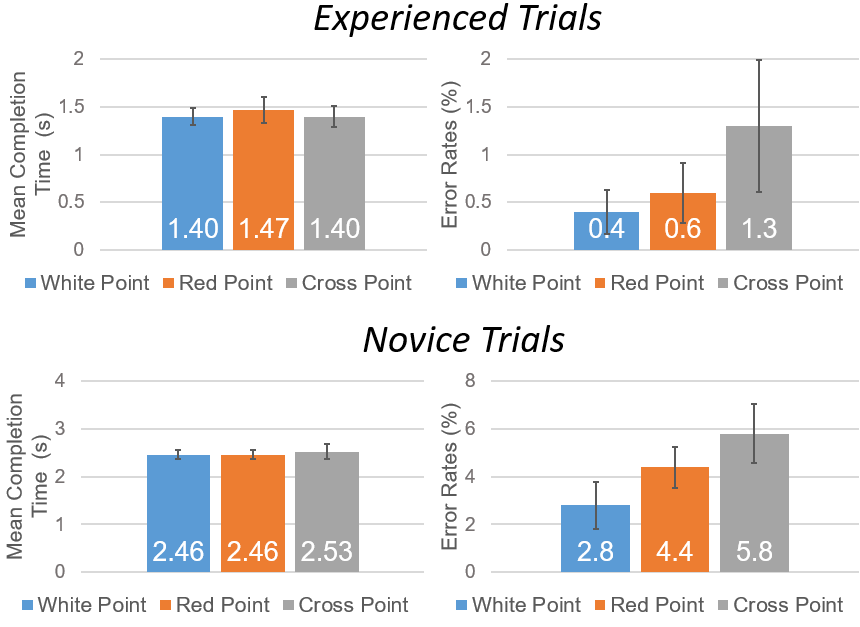}
  \caption{Mean of CT (s) and ER (\%) from preliminary study 2. Error bars indicate standard errors. }
  \label{fig:preliminary2result}
  \Description{Mean of CT and ER from preliminary study 2. Error bars indicate standard errors.}
\end{figure}

For both \textit{Experienced Trials} and \textit{Novice Trials}, the effects of \textit{VisualAnchorType} on both CT and ER were not significant.

\subsubsection{Post-interview on user experience}

In the post-interview, most of the subjects (eight) chose the \textit{White Point} as their preferred condition. For the most disliked condition, a majority of the subjects (nine) chose the \textit{Red Point}. They mainly commented on the drawbacks of "over-saliency," stating that \textit{"It causes unintended gaze shifts"} (p1), \textit{"it was too bright"} (p2 and p6), and \textit{"I felt higher eye fatigue."} (p5, p7, p8, p11, and p12). Similarly for the \textit{Cross Point}, the subjects commented on their eye fatigue (p3 and p12). In addition, p6 and p11 described that the multiple \textit{Cross Points} on the screen made them feel dizzy.

We decided to use the \textit{White Point} for our final design. Even considering that the \textit{White Point} might be favorable because the subjects were already familiarized with it from preliminary study 1, negative user experiences regarding the \textit{Red Point} and the \textit{Cross Point} were crucial. Moreover, we also considered that the \textit{White Point} showed the lowest error rates for both \textit{Experienced Trials} and \textit{Novice Trials} though the differences were not statistically significant. 

\section{User Study 1: Performance Evaluation}

\begin{figure}[b]
  \centering
  \includegraphics[width=6cm]{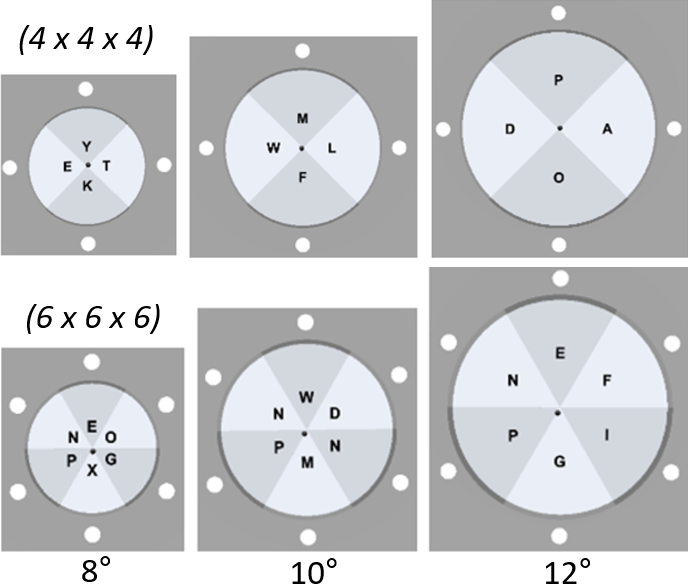}
  \caption{Various menu layouts tested in User Study 1.}
  \label{fig:userstudy1layouts}
  \Description{Various menu layouts tested in User Study 1 including three menu sizes and two menu structures}
\end{figure}

In User Study 1, we formally evaluated the performance of the designed Lattice Menu on various menu layouts with different menu sizes and structures. \textit{Structure} implies the number of items in each menu level. The experiment was a two-way within-subject design with the following independent variables and levels. The order of the conditions was counterbalanced using Balanced Latin Square.

\begin{itemize}
  \item Structure: \textit{4 $\times$ 4 $\times$ 4} and \textit{6 $\times$ 6 $\times$ 6}
  \item Size (\textit{D3}, Effective Radius): \textit{8}, \textit{10}, and \textit{12\textdegree{}}
\end{itemize}

\subsection{Procedure}

For each \textit{Structure} $\times$ \textit{Size} condition, the same six target paths (two each from \textit{0-bent}, \textit{1-bent}, and \textit{2-bent} chosen at random) were tested. One block consisted of 144 trials (2 \textit{Structures} $\times$ 3 \textit{Size} $\times$ 6 paths $\times$ 4 repetitions) of the menu selection task and the subject performed five blocks without prior training because they were already trained after 3 days of experience on the interfaces. It took approximately 1.5 h to complete the experiment. 

\subsection{Analysis}

We collected 2160 \textit{Novice Trials} (2 \textit{Structures} $\times$ 3 \textit{Sizes} $\times$ 6 paths $\times$ 5 blocks $\times$ 12 subjects) and 4320 \textit{Experienced Trials}, and calculated the CT and ER. For the analysis, we performed a two-way ANOVA on CT and ER. We applied an aligned rank transform (ART) \cite{wobbrock2011aligned} on ER before conducting RM-ANOVA. For a post hoc comparison, a paired sample t-test on CT and a Wilcoxon signed-rank test on ER with a Bonferroni correction was used.

\subsection{Results}

\begin{figure}[t]
  \centering
  \includegraphics[width=8.5cm]{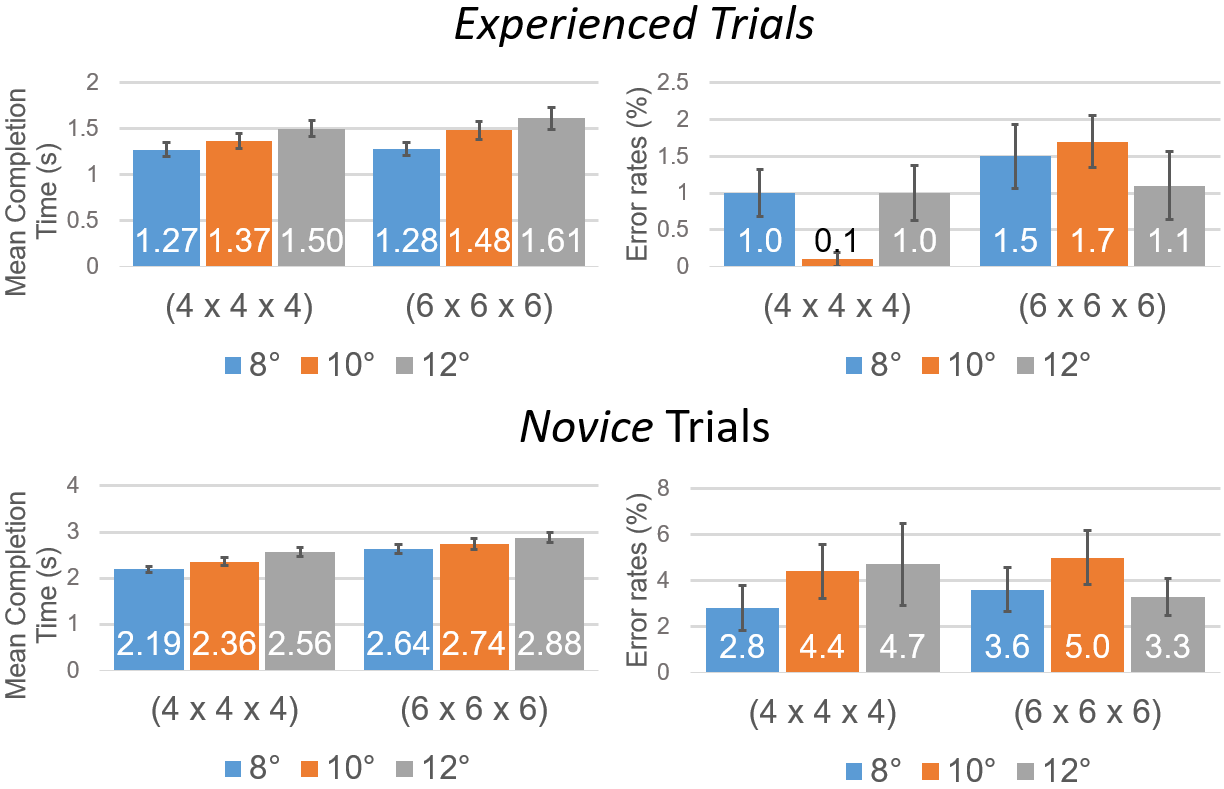}
  \caption{Mean of CT (s) and ER (\%) from User Study 1. Error bars indicate standard errors. }
  \label{fig:userstudy1result}
  \Description{Mean of CT and ER from User Study 1. Error bars indicate standard errors.}
\end{figure}

\subsubsection{Experienced Trials}

In the case of CT, the effects of both \textit{Structure} (\textit{F}(1,11) = 10.964, \textit{p} < .01) and \textit{Size} (\textit{F}(2,22) = 19.070, \textit{p} < .001) were significant. No significant interaction effects were observed. The post hoc comparison under each factor revealed that there were significant differences between \textit{Size} conditions (8\textdegree{} vs. 10\textdegree{}, \textit{t} = -3.530, \textit{p} < .05; 8\textdegree{} vs. 12\textdegree{}, \textit{t} = -5.396, \textit{p} < .005; 10\textdegree{} vs. 12\textdegree{}, \textit{t} = -3.181, \textit{p} < .05) and \textit{Structure} conditions (4 $\times$ 4 $\times$ 4 vs. 6 $\times$ 6 $\times$ 6, \textit{t} = -3.311, \textit{p} < .05). In the case of ER, the effect of \textit{Structure} was significant (\textit{F}(1,11) = 4.468, \textit{p} < .05) whereas the effect of \textit{Size} was not. There was a significant interaction effect (\textit{F}(2,22) = 3.538, \textit{p} < .05).

\subsubsection{Novice Trials}

In the case of CT, the effects of both \textit{Structure} (\textit{F}(1,11) = 86.973, \textit{p} < .001) and \textit{Size} (\textit{F}(2,22) = 16.893, \textit{p} < .001) were significant. No significant interaction effects were observed. The post hoc comparison under each factor revealed that there were significant differences between \textit{Size} conditions (8\textdegree{} vs. 12\textdegree{}, \textit{t} = -8.361, \textit{p} < .005; 10\textdegree{} vs. 12\textdegree{}, \textit{t} = -3.228, \textit{p} < .05) and \textit{Structure} conditions (4 $\times$ 4 $\times$ 4 vs. 6 $\times$ 6 $\times$ 6, \textit{t} = -8.261, \textit{p} < .005). In the case of ER, the effects of both \textit{Structure} and \textit{Size} were not significant.

\subsection{Discussion}

For the menu selection time in expert usage (\textit{Experienced Trials}), we observed similar performance between 4 $\times$ 4 $\times$ 4 and 6 $\times$ 6 $\times$ 6 (1.38 vs. 1.46 s on average) which implies that the target-assisted gaze gesture with a diagonal eye movement can be easily performed as well. We can observe general increase of menu selection time with an increase in the number of items (\textit{Structure}) and menu sizes (\textit{Size}). For \textit{Experienced Trials}, the effect of \textit{Size} was dominant, whereas for \textit{Novice Trials}, the effect of the number of items (\textit{Structure}) was dominant. For the selection errors, 6 $\times$ 6 $\times$ 6 showed generally higher error rates than the 4 $\times$ 4 $\times$ 4 \textit{Structure} in both novice and expert usage.

For overall performance, Lattice Menu showed considerably low error rates (0.7\% for \textit{4 $\times$ 4 $\times$ 4} and 1.4\% for \textit{6 $\times$ 6 $\times$ 6} \textit{Structures}) and short menu selection time (1.3-1.6 s) for expert usage (i.e., \textit{Experienced Trials}). For the novice usage (i.e., \textit{Novice Trials}), the menu item selection took 2.2-2.9 s with an overall error rate of \textasciitilde4\%. 

\section{User Study 2: Comparing Lattice Menu with BorderPie}

In User Study 2, we compared the performance of Lattice Menu with the traditional border-crossing marking menu design \cite{urbina2010pies} (\textit{BorderPie}) that does not utilize visual targets. The purpose of this study is to investigate the performance improvement of Lattice Menu over traditional technique.

The experiment was a two-way within-subject design with the following independent variables and levels. The order of the \textit{Technique} $\times$ \textit{Structure} conditions was counterbalanced using Balanced Latin Square.

\begin{itemize}
  \item Technique: \textit{Lattice Menu}, Border-crossing Marking Menu (\textit{BorderPie})
  \item Structure: 4 $\times$ 4 $\times$ 4, 6 $\times$ 6 $\times$ 6
\end{itemize}

\subsection{Procedure}

The experiment was conducted over 4 consecutive days. Subjects experienced different \textit{Technique} $\times$ \textit{Structure} conditions each day. Three menu sizes (\textit{D3}: 8, 10, and 12\textdegree{}) were tested. For each \textit{Technique} $\times$ \textit{Structure} $\times$ \textit{Size} condition, the same 16 target paths were provided; 1, 6, and 9 paths from \textit{0-bent}, \textit{1-bent}, and \textit{2-bent} were randomly picked for 4 $\times$ 4 $\times$ 4 \textit{Structure}, and 1, 4, and 11 paths from \textit{0-bent}, \textit{1-bent}, and \textit{2-bent} were randomly picked for 6 $\times$ 6 $\times$ 6 \textit{Structure} considering the ratio of paths for each \textit{Structure}) tested. The experiment each day consisted of 192 trials (3 \textit{Sizes} $\times$ 16 paths $\times$ 4 repetitions) for the training and the same number of trials for the testing. It took less than 1 h for a participant to complete the experiment each day.

\subsection{Analysis}

We collected 2304 \textit{Novice Trials} (2 \textit{Techniques} $\times$ 2 \textit{Structures} $\times$ 3 menu sizes $\times$ 16 paths $\times$ 12 subjects) and 4608 \textit{Experienced Trials}, and calculated the CT and ER. For the analysis, we performed a two-way ANOVA on CT and ER. We applied an aligned rank transform (ART) to ER before conducting the RM-ANOVA. For a post hoc comparison, a paired sample t-test on CT and a Wilcoxon signed-rank test on ER with a Bonferroni correction were conducted.

\subsection{Results}

\begin{figure}[t]
  \centering
  \includegraphics[width=8.5cm]{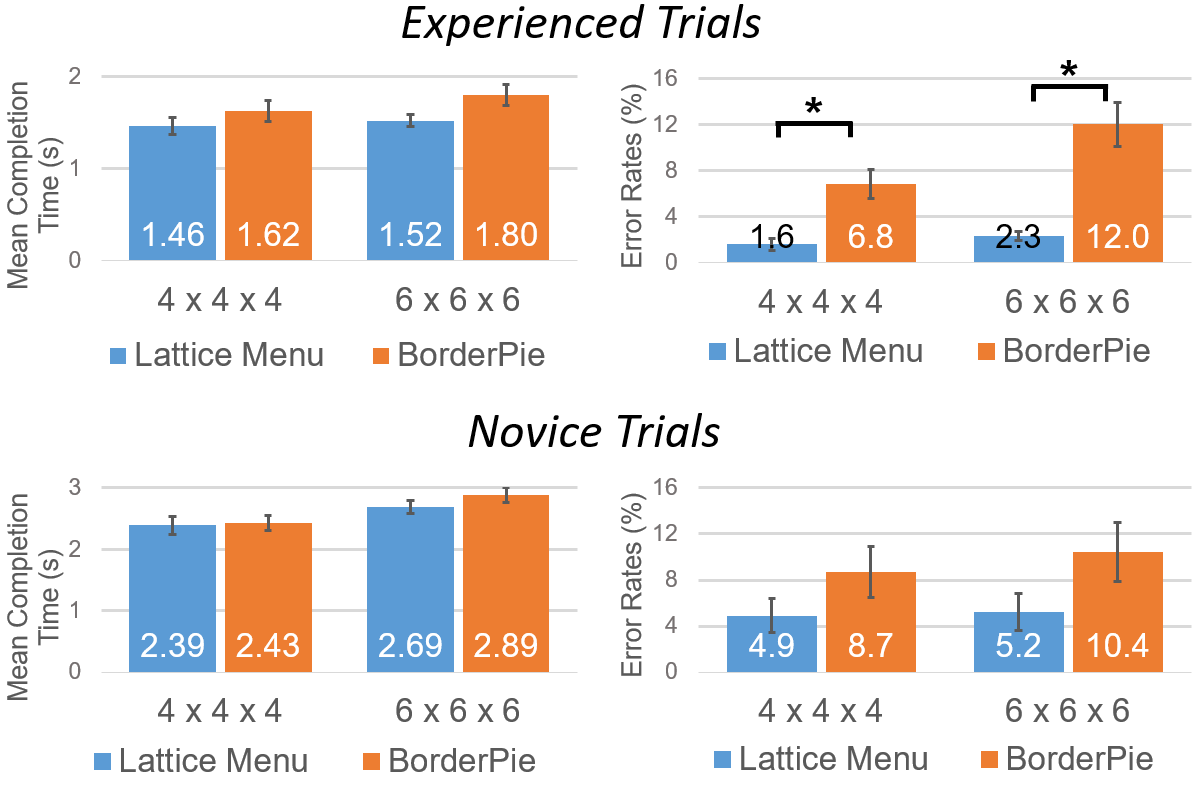}
\caption{Mean of CT (s) and ER (\%) from User Study 2. Error bars show standard errors, and asterisks indicate significant differences (\textit{p} < .05). }  \label{fig:userstudy2result}
  \Description{Mean of CT and ER from User Study 2. Error bars show standard errors, and asterisks indicate significant differences.}
\end{figure}

\subsubsection{Experienced Trials}

In the case of CT, the effects of both \textit{Technique} and \textit{Structure} were not significant. In the case of ER, the effects of both \textit{Technique} (\textit{F}(1,11) = 53.400, \textit{p} < .001) and \textit{Structure} (\textit{F}(1,11) = 14.458, \textit{p} < .001) were significant. There was a significant interaction effect (\textit{F}(1,11) = 7.446, \textit{p} < .05). The post hoc comparison revealed that there were significant difference between pairs (4 $\times$ 4 $\times$ 4 \textit{Lattice Menu} vs. 4 $\times$ 4 $\times$ 4 \textit{BorderPie}, \textit{Z} = -2.703, \textit{p} <.05; 6 $\times$ 6 $\times$ 6 \textit{Lattice Menu} vs. 6 $\times$ 6 $\times$ 6 \textit{BorderPie}, \textit{Z} = -2.934, \textit{p} <.05).

\subsubsection{Novice Trials}

In the case of CT, the effect of \textit{Technique} was not significant; however, the effect of \textit{Structure} (\textit{F}(1,11) = 56.855, \textit{p} < .001) was significant. No significant interaction effects were observed. In the case of ER, the effect of \textit{Technique} was significant (\textit{F}(1,11) = 7.060, \textit{p} < .05), whereas the effect of \textit{Structure} was not significant. No significant interaction effects were observed.

\subsection{Discussion}

\begin{figure}[t]
  \centering
  \includegraphics[width=8.5cm]{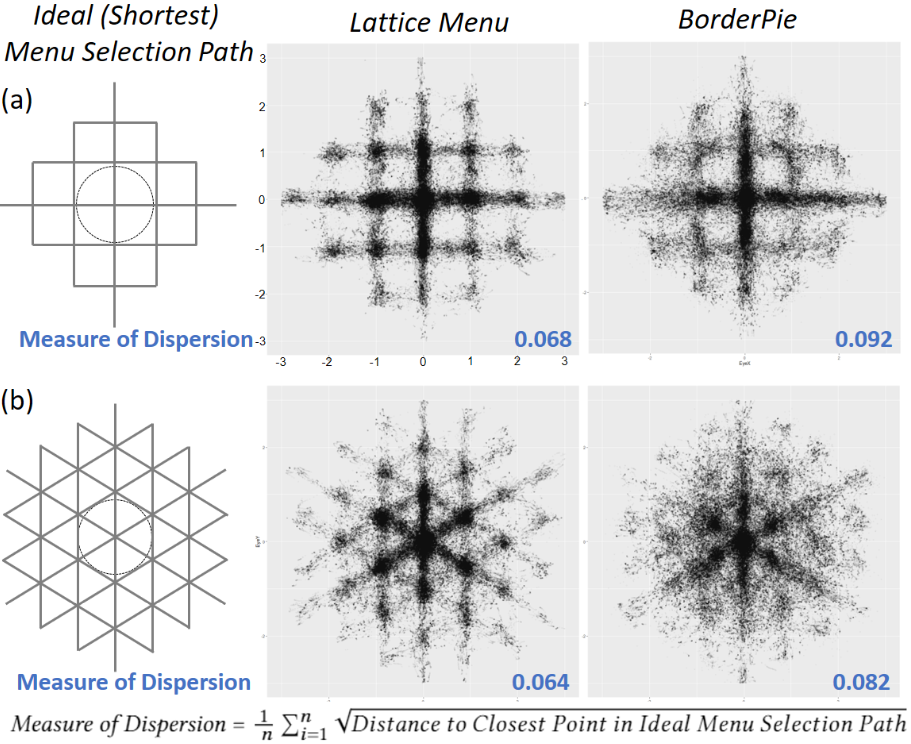}
  \caption{ Distribution of saccade landing position for (a) 4 $\times$ 4 $\times$ 4 and (b) 6 $\times$ 6 $\times$ 6 \textit{Structures} (normalized scale, n $\approx$ 300,000 for each plot) from User Study 2 and calculated \textit{Measure of Dispersion} with the illustration of ideal menu selection paths.}
  \label{fig:userstudy2visualize}
  \Description{ Distribution of saccade landing position from User Study 2 and calculated Measure of Dispersion with the illustration of ideal menu selection paths.}
\end{figure}

The comparative evaluation showed the remarkably (\textasciitilde5 times) lower error rates of \textit{Lattice Menu} in comparison to \textit{BorderPie} for expert usage (i.e., \textit{Experienced Trials}). \textit{BorderPie} in particular showed a steep increase in error rate in a 6 $\times$ 6 $\times$ 6 \textit{Structure} (12\%). This observation implies that visual targets can be more beneficial when the number of items in each menu level increases. In general, there were no significant differences in the menu selection time; however, there was a remarkable difference in the error rates. We observed approximately 2 (\textit{Novice Trials}) to 5 (\textit{Experienced Trials}) times higher error rates in \textit{BorderPie}. The scatter plot of the saccade landing positions shown in Figure \ref{fig:userstudy2visualize} also indicates the effectiveness of the \textit{Lattice Menu} technique.  The lattice of visual anchors allowed the users to better follow the ideal paths for selection.

In the post-interview on user experience, all 12 subjects preferred to use \textit{Lattice Menu} over \textit{BorderPie}. Most of the subjects (eight) noted that they felt more tired with increased eye fatigue when using \textit{BorderPie}. The main comments on \textit{Lattice Menu} are as follows: (1) Having a visual target to focus helps accurate gaze pointing (\textit{p1, p4, p5, p7, p9}). (2) There was not much difference in difficulty between 4 $\times$ 4 $\times$ 4 vs. 6 $\times$ 6 $\times$ 6 \textit{Lattice Menu} (\textit{p5, p6}), and (3) \textit{Lattice Menu} was more sensitive to the tracking offsets because of the smaller area used for selection (\textit{p4, p5}). The main comments on \textit{BorderPie} are as follows: (1) Errors often occur by moving the eyes more than intended (\textit{p1, p10}). (2) As the menu size becomes smaller, unintended selections occurs more frequently (\textit{p2, p8, p12}). These observed types of errors in \textit{BorderPie} are consistent with the observations in StickyPie \cite{ahn2021stickypie}.

\section{General Discussion and Future Work}

\subsection{Design Insight for Gaze-Based Menu}

We share several insights for gaze-based menu design based on the observations in this study.

\textbf{(1) Visual anchors effectively guide the eye gaze for safe menu selection.} Several unwanted options are always located adjacent to a single wanted item. Because of the ballistic nature of human eye movement \cite{majaranta2019eye}, designing a safe gaze-based menu selection is challenging. In this study, we confirmed that the provisioning of visual anchors can safely guide eye movement and prevent unintentional selection errors in the gaze-based menu.

\textbf{(2) Over-saliency for the visual guidance target results in poor user experiences. } The visual target for eye guidance should be sufficiently salient to encourage effortless \cite{le2015saccadic} and accurate \cite{deuble1984evaluation} gaze pointing. However, as observed in preliminary study 2, excessive saliency can induce increased eye fatigue and impact the overall user experience at the interface. The target saliency must be carefully designed for comfortable eye guidance.

\textbf{(3) The space for the menu contents (e.g., label texts or icons) needs to be separated from the space for the menu item selection.} Separation of the two spaces allows users to navigate the menu items safely. When two spaces are not separated, unintentional selection errors easily occur as observed in the pilot study noted in Section 3.2. We suggest making an explicit boundary to capture the eye \cite{theeuwes1998our} inside the safe region during user menu navigation.

\subsection{Applying Lattice Menu in Real Application}

\subsubsection{External Validity: Surrounding Environment}

We used a plain gray background for the controlled experimental setup in this study. However, in a real application, the surrounding AR/VR environment and UI contents can be dynamic, which should be considered when adopting Lattice Menu. We recommend showing Lattice Menu under an opaque background if the application context allows the background to be occluded. Giving a border to the visual anchor circle \cite{majaranta2019inducing} may also be helpful for distinguishing the anchor from the background.

\subsubsection{External Validity: Size and Depth}

In this study, we used a menu size (\textit{D3} in Figure \ref{fig:latticeMenuParameters}) of between 8\textdegree{} and 12\textdegree{} with a three-level menu depth. However, depending on the target situation, designers may need to limit the size or depth of Lattice Menu. Reducing the menu depth to two-level can be effective if the required menu capacity of the target application is small (e.g., using 4 $\times$ 4 Lattice Menu when required number of commands is less than 16). \textit{D3} and \textit{D4} may be reduced until the space limit for the menu contents (e.g., icon or text). Using symbolic icons instead of long texts can be a good option. Designers can also adjust \textit{D2} considering the tracking offset of the eye tracker. 

\subsubsection{Design Opportunity: Personalizing Lattice Menu}

Designers can change the parameter values (depicted in Figure \ref{fig:latticeMenuParameters}) to personalize Lattice Menu. Because there are large individual differences in ocular motor control skills and eye tracking offsets, personalizing Lattice Menu is expected to be beneficial. Decreasing \textit{D4} (i.e., making the visual pie smaller) will be helpful for users who suffer from unintentional selection errors during menu navigation because it will provide the distance between the visual pie and \textit{Item Selection Zone}. In addition, increasing \textit{D2} to greater than 2\textdegree{} will be helpful for users who have trouble selecting items owing to the large eye tracking offsets. 

\subsection{Head-Fixed Lattice Menu}

As we described in Section 3.1, there can be two types of menus in the AR/VR environment; \textit{Head-Fixed Menu} that uses eyes-only input and \textit{World-Fixed Menu} which uses a head-combined gaze input for control. The head-combined gaze input can be comfortable \cite{sidenmark2019eye} for gaze shifts of over 25\textdegree{}, whereas eyes-only input has a more limited range of comfortable shifts (\textasciitilde20\textdegree{} \cite{sidenmark2019eyehead}). Designers can properly utilize each type of menu considering the target use case in the AR/VR. (e.g., using \textit{Head-Fixed Menu} for a global-level menu and \textit{World-Fixed Menu} for a context menu with a certain target object) 

In this study, Lattice Menu was implemented and evaluated using a head-combined gaze input, i.e., a \textit{World-Fixed Menu} setup. Validation in this setup has the advantage of being compatible with an external, stationary eye tracker (e.g., for smart TV) environment because head-combined gaze input is utilized in both cases. Meanwhile, validation in the eyes-only input (\textit{Head-Fixed Menu}) setup also has its advantage, being the only possible option when 3D sensing for the surrounding space is incapable of (e.g., smart glasses without 3D sensing capability).

To design Head-Fixed Lattice Menu with eyes-only control, further investigation (e.g., a usable menu size and hierarchy) is necessary considering the differences in motor skills between eyes-only and head-combined eye movements. Eyes-only, Head-Fixed Lattice Menu will be further explored in our future study.

\subsection{Comparison with StickyPie}

\begin{table}[t]
  \caption{The comparison between StickyPie \cite{ahn2021stickypie} and Lattice Menu}
  \label{tab:commands}
  \begin{tabular} {|c|c|c|}
    \hline
         & Lattice Menu & StickyPie \\
    \hline
    Tested Input & \begin{tabular}{@{}c@{}} Head-Combined \\ Gaze Input\end{tabular} & Eyes-only Input \\
    \hline
    \begin{tabular}{@{}c@{}}Reported Menu \\ Selection Time\end{tabular} & 1.4-2.6s & 2.5-3.4s \\
    \hline
    \begin{tabular}{@{}c@{}}Reported Menu \\ Selection Error Rate\end{tabular} &  2-6\% & \textasciitilde15\%  \\
    \hline
    \end{tabular}
    \Description{The comparison between StickyPie and the Lattice Menu}
\end{table}

Similar to this work, StickyPie \cite{ahn2021stickypie} proposes a way to make the gaze-based marking menu less error-prone. StickyPie predicts the saccade landing position to prevent overshooting errors, whereas Lattice Menu provides visual anchors for stable eye control. Because StickyPie and Lattice Menu pursue different ideas for the same goal, a comparison between the two should be made. However, as described in Section 2.2, a direct comparison was not conducted in this study because of practical difficulties. 

% BorderPie: 32-64

Instead, we attempted to compare the reported performances of each technique. The 6 $\times$ 6 $\times$ 6 Lattice Menu with an 8\textdegree{} radius condition tested in User Study 2 of this paper has a similar menu layout with the \textit{G6} condition (6 $\times$ 6 $\times$ 6 menu structure with a 7.5\textdegree{} radius) in Experiment 2 of the StickyPie study. Both conditions underwent a similar level of training (32-64 repetitions). For \textit{Experienced Trial}, menu selection on \textit{Lattice Menu} (1.4s, 2\%) was 1.8-times faster and showed almost 8-times lower error rates than \textit{StickyPie} (2.5 s, ~15\%). For \textit{Novice Trial}, menu selection on \textit{Lattice Menu} (2.6s, ~6\%) was approximately 1.3 times faster and showed 3 times less error rates than \textit{StickyPie} (3.4s, ~15\%). The results of the comparisons are summarized in Table 1. Though the difference in tested input need to be considered, we were able to observed remarkably lower error rates in Lattice Menu from the reported performances.

\subsection{Cancel and Back Menu Operation}

Although not explored in this study, \textit{Cancel} (i.e., closing a menu without selection) and \textit{Back} (i.e., a roll-back to the previous menu-level) are important functionalities for the menu. For \textit{Cancel}, we considered utilizing an eye behavior that does not naturally occur during the menu selection process (e.g., long eye winks/blinks). For \textit{Back}, we considered reserving one menu item in each submenu for the roll-back operation. With this approach, the 4 $\times$ 4 $\times$ 4 layout can cover 36 commands (4 $\times$ 3 $\times$ 3), and the 6 $\times$ 6 $\times$ 6 layout can cover 150 commands (6 $\times$ 5 $\times$ 5). 

\section{Conclusion}

This study introduced Lattice menu, a gaze-based marking menu utilizing target-assisted gaze gestures on a lattice of visual anchors. From our empirical evaluation, Lattice Menu showed a considerably low error rate (\textasciitilde1\%) and quick menu selection time (1.3-1.6 s) on various menu layouts. The distribution of the saccade landing position demonstrated the effectiveness of the visual anchors on stable gaze guidance. We also shared several design insights for the gaze-based menu technique from the observed user behaviors. We hope this study can be a meaningful step on realizing practical hands-free menu interaction.

\begin{acks}
This work was supported by Institute for Information \& Communications Technology Promotion (IITP) grant funded by the Korea government (MSIT) (No.2020-0-00537, Development of 5G based low latency device – edge cloud interaction technology).
\end{acks}

\bibliographystyle{ACM-Reference-Format}
\bibliography{main}

@inproceedings{urbina2010pies,
  title={Pies with EYEs: the limits of hierarchical pie menus in gaze control},
  author={Urbina, Mario H and Lorenz, Maike and Huckauf, Anke},
  booktitle={Proceedings of the 2010 Symposium on Eye-Tracking Research \& Applications},
  pages={93--96},
  year={2010}
}

@book{kurtenbach1993design,
  title={The design and evaluation of marking menus.},
  author={Kurtenbach, Gordon Paul},
  year={1993},
  publisher={University of Toronto Toronto}
}

@inproceedings{kurtenbach1991issues,
  title={Issues in combining marking and direct manipulation techniques},
  author={Kurtenbach, Gordon and Buxton, William},
  booktitle={Proceedings of the 4th annual ACM symposium on User interface software and technology},
  pages={137--144},
  year={1991}
}

@inproceedings{kurtenbach1994user,
  title={User learning and performance with marking menus},
  author={Kurtenbach, Gordon and Buxton, William},
  booktitle={Proceedings of the SIGCHI conference on Human factors in computing systems},
  pages={258--264},
  year={1994}
}

@article{hornof2003eyedraw,
  title={Eyedraw: a system for drawing pictures with eye movements},
  author={Hornof, Anthony and Cavender, Anna and Hoselton, Rob},
  journal={ACM SIGACCESS Accessibility and Computing},
  number={77-78},
  pages={86--93},
  year={2003},
  publisher={ACM New York, NY, USA}
}

@inproceedings{wobbrock2008longitudinal,
  title={Longitudinal evaluation of discrete consecutive gaze gestures for text entry},
  author={Wobbrock, Jacob O and Rubinstein, James and Sawyer, Michael W and Duchowski, Andrew T},
  booktitle={Proceedings of the 2008 symposium on Eye tracking research \& applications},
  pages={11--18},
  year={2008}
}

@inproceedings{majaranta2019inducing,
  title={Inducing gaze gestures by static illustrations},
  author={Majaranta, P{\"a}ivi and Laitinen, Jari and Kangas, Jari and Isokoski, Poika},
  booktitle={Proceedings of the 11th ACM Symposium on Eye Tracking Research \& Applications},
  pages={1--5},
  year={2019}
}

@inproceedings{drewes2007interacting,
  title={Interacting with the computer using gaze gestures},
  author={Drewes, Heiko and Schmidt, Albrecht},
  booktitle={IFIP Conference on Human-Computer Interaction},
  pages={475--488},
  year={2007},
  organization={Springer}
}

@inproceedings{henderson2020investigating,
  title={Investigating the necessity of delay in marking menu invocation},
  author={Henderson, Jay and Malacria, Sylvain and Nancel, Mathieu and Lank, Edward},
  booktitle={Proceedings of the 2020 CHI Conference on Human Factors in Computing Systems},
  pages={1--13},
  year={2020}
}

@inproceedings{ahn2021stickypie,
  title={StickyPie: A Gaze-Based, Scale-Invariant Marking Menu Optimized for AR/VR},
  author={Ahn, Sunggeun and Santosa, Stephanie and Parent, Mark and Wigdor, Daniel and Grossman, Tovi and Giordano, Marcello},
  booktitle={Proceedings of the 2021 CHI Conference on Human Factors in Computing Systems},
  pages={1--16},
  year={2021}
}

@inproceedings{kammerer2008looking,
  title={Looking my way through the menu: the impact of menu design and multimodal input on gaze-based menu selection},
  author={Kammerer, Yvonne and Scheiter, Katharina and Beinhauer, Wolfgang},
  booktitle={Proceedings of the 2008 Symposium on Eye Tracking Research \& Applications},
  pages={213--220},
  year={2008}
}

@inproceedings{huckauf2008gazing,
  title={Gazing with pEYEs: towards a universal input for various applications},
  author={Huckauf, Anke and Urbina, Mario H},
  booktitle={Proceedings of the 2008 symposium on Eye tracking research \& applications},
  pages={51--54},
  year={2008}
}

@article{isomoto2020gaze,
  title={Gaze-based Command Activation Technique Robust Against Unintentional Activation using Dwell-then-Gesture},
  author={Isomoto, Toshiya and Yamanaka, Shota and Shizuki, Buntarou},
  year={2020}
}

@inproceedings{hyrskykari2012gaze,
  title={Gaze gestures or dwell-based interaction?},
  author={Hyrskykari, Aulikki and Istance, Howell and Vickers, Stephen},
  booktitle={Proceedings of the Symposium on Eye Tracking Research and Applications},
  pages={229--232},
  year={2012}
}

@article{thaler2013best,
  title={What is the best fixation target? The effect of target shape on stability of fixational eye movements},
  author={Thaler, Lore and Sch{\"u}tz, Alexander C and Goodale, Melvyn A and Gegenfurtner, Karl R},
  journal={Vision research},
  volume={76},
  pages={31--42},
  year={2013},
  publisher={Elsevier}
}

@article{menozzi1994direction,
  title={Direction of gaze and comfort: discovering the relation for the ergonomic optimization of visual tasks},
  author={Menozzi, M v and Buol, A v and Krueger, H and Mi{\`e}ge, Ch},
  journal={Ophthalmic and Physiological Optics},
  volume={14},
  number={4},
  pages={393--399},
  year={1994},
  publisher={Wiley Online Library}
}

@article{sidenmark2019eye,
  title={Eye, head and torso coordination during gaze shifts in virtual reality},
  author={Sidenmark, Ludwig and Gellersen, Hans},
  journal={ACM Transactions on Computer-Human Interaction (TOCHI)},
  volume={27},
  number={1},
  pages={1--40},
  year={2019},
  publisher={ACM New York, NY, USA}
}

@article{theeuwes2010object,
  title={Object-based eye movements: The eyes prefer to stay within the same object},
  author={Theeuwes, Jan and Math{\^o}t, Sebastiaan and Kingstone, Alan},
  journal={Attention, Perception, \& Psychophysics},
  volume={72},
  number={3},
  pages={597--601},
  year={2010},
  publisher={Springer}
}

@article{le2015saccadic,
  title={Saccadic model of eye movements for free-viewing condition},
  author={Le Meur, Olivier and Liu, Zhi},
  journal={Vision research},
  volume={116},
  pages={152--164},
  year={2015},
  publisher={Elsevier}
}

@incollection{deuble1984evaluation,
  title={The evaluation of the oculomotor error signal},
  author={Deuble, H and Wolf, W and Hauske, G},
  booktitle={Advances in Psychology},
  volume={22},
  pages={55--62},
  year={1984},
  publisher={Elsevier}
}

@article{kalesnykas1994retinal,
  title={Retinal eccentricity and the latency of eye saccades},
  author={Kalesnykas, RP and Hallett, PE},
  journal={Vision research},
  volume={34},
  number={4},
  pages={517--531},
  year={1994},
  publisher={Elsevier}
}

@article{bell2006stimulus,
  title={Stimulus intensity modifies saccadic reaction time and visual response latency in the superior colliculus},
  author={Bell, AH and Meredith, MA and Van Opstal, AJ and Munoz, DougP},
  journal={Experimental Brain Research},
  volume={174},
  number={1},
  pages={53--59},
  year={2006},
  publisher={Springer}
}

@article{itti2001computational,
  title={Computational modelling of visual attention},
  author={Itti, Laurent and Koch, Christof},
  journal={Nature reviews neuroscience},
  volume={2},
  number={3},
  pages={194--203},
  year={2001},
  publisher={Nature Publishing Group}
}

@inproceedings{schuetz2019explanation,
  title={An Explanation of Fitts' Law-like Performance in Gaze-Based Selection Tasks Using a Psychophysics Approach},
  author={Schuetz, Immo and Murdison, T Scott and MacKenzie, Kevin J and Zannoli, Marina},
  booktitle={Proceedings of the 2019 CHI Conference on Human Factors in Computing Systems},
  pages={1--13},
  year={2019}
}

@article{bhattarai2019fixation,
  title={Fixation stability with Bessel beams},
  author={Bhattarai, Dipesh and Suheimat, Marwan and Lambert, Andrew J and Atchison, David A},
  journal={Optometry and Vision Science},
  volume={96},
  number={2},
  pages={95--102},
  year={2019},
  publisher={LWW}
}

@inproceedings{wobbrock2011aligned,
  title={The aligned rank transform for nonparametric factorial analyses using only anova procedures},
  author={Wobbrock, Jacob O and Findlater, Leah and Gergle, Darren and Higgins, James J},
  booktitle={Proceedings of the SIGCHI conference on human factors in computing systems},
  pages={143--146},
  year={2011}
}

@inproceedings{sidenmark2019eyehead,
  title={Eye\&head: Synergetic eye and head movement for gaze pointing and selection},
  author={Sidenmark, Ludwig and Gellersen, Hans},
  booktitle={Proceedings of the 32nd Annual ACM Symposium on User Interface Software and Technology},
  pages={1161--1174},
  year={2019}
}

@Misc{FOVE,
  author = {FOVE inc.},
  year = {2021},
  title =        {FOVE0 Headset Specification},
  note = {\url{https://fove-inc.com/product/}, last visited Dec. 2021}
}

@Misc{ViveProEye,
  author = {HTC inc.},
  year = {2021},
  title =        {HTC Vive Pro Eye Headset Specification},
  note = {\url{https://www.vive.com/kr/product/vive-pro-eye/specs/}, last visited Dec. 2021}
}

@inproceedings{majaranta2009fast,
  title={Fast gaze typing with an adjustable dwell time},
  author={Majaranta, P{\"a}ivi and Ahola, Ulla-Kaija and {\v{S}}pakov, Oleg},
  booktitle={Proceedings of the SIGCHI Conference on Human Factors in Computing Systems},
  pages={357--360},
  year={2009}
}

@inproceedings{tien2008improving,
  title={Improving hands-free menu selection using eyegaze glances and fixations},
  author={Tien, Geoffrey and Atkins, M Stella},
  booktitle={Proceedings of the 2008 symposium on Eye tracking research \& applications},
  pages={47--50},
  year={2008}
}

@inproceedings{scarr2011dips,
  title={Dips and ceilings: understanding and supporting transitions to expertise in user interfaces},
  author={Scarr, Joey and Cockburn, Andy and Gutwin, Carl and Quinn, Philip},
  booktitle={Proceedings of the sigchi conference on human factors in computing systems},
  pages={2741--2750},
  year={2011}
}

@article{cockburn2014supporting,
  title={Supporting novice to expert transitions in user interfaces},
  author={Cockburn, Andy and Gutwin, Carl and Scarr, Joey and Malacria, Sylvain},
  journal={ACM Computing Surveys (CSUR)},
  volume={47},
  number={2},
  pages={1--36},
  year={2014},
  publisher={ACM New York, NY, USA}
}

@article{theeuwes1998our,
  title={Our eyes do not always go where we want them to go: Capture of the eyes by new objects},
  author={Theeuwes, Jan and Kramer, Arthur F and Hahn, Sowon and Irwin, David E},
  journal={Psychological Science},
  volume={9},
  number={5},
  pages={379--385},
  year={1998},
  publisher={SAGE Publications Sage CA: Los Angeles, CA}
}

@article{yantis1994stimulus,
  title={Stimulus-driven attentional capture: evidence from equiluminant visual objects.},
  author={Yantis, Steven and Hillstrom, Anne P},
  journal={Journal of experimental psychology: Human perception and performance},
  volume={20},
  number={1},
  pages={95},
  year={1994},
  publisher={American Psychological Association}
}

@inproceedings{porta2008eye,
  title={Eye-S: a full-screen input modality for pure eye-based communication},
  author={Porta, Marco and Turina, Matteo},
  booktitle={Proceedings of the 2008 symposium on Eye tracking research \& applications},
  pages={27--34},
  year={2008}
}

@inproceedings{rajanna2018gaze,
  title={Gaze typing in virtual reality: impact of keyboard design, selection method, and motion},
  author={Rajanna, Vijay and Hansen, John Paulin},
  booktitle={Proceedings of the 2018 ACM Symposium on Eye Tracking Research \& Applications},
  pages={1--10},
  year={2018}
}

@incollection{majaranta2019eye,
  title={Eye movements and human-computer interaction},
  author={Majaranta, P{\"a}ivi and R{\"a}ih{\"a}, Kari-Jouko and Hyrskykari, Aulikki and {\v{S}}pakov, Oleg},
  booktitle={Eye Movement Research},
  pages={971--1015},
  year={2019},
  publisher={Springer}
}

\appendix

\end{document}